\documentclass[aps,amssymb,prd]{revtex4}

\usepackage{appendix}
\usepackage{amsbsy}
\usepackage{amsfonts}
\usepackage{amsmath}
\usepackage{amssymb}
\usepackage{amsthm}
\usepackage{xcolor}

\usepackage{graphicx}
\usepackage{ifthen}
\usepackage{textcomp}

\newcounter{multi} \newcounter{multa}

\newcounter{faki} \newcounter{faka}

\theoremstyle{definition} 

\theoremstyle{remark} 

\newcommand{\beqa}{\begin{eqnarray}}
\newcommand{\beq}{\begin{equation}}
\newcommand{\eeqa}{\end{eqnarray}}
\newcommand{\eeq}{\end{equation}}

\newcommand{\fkM}{\mathfrak{\mu}}

\def\RR{\mathbb{R}}

\begin{document}

\title{Magnification Cross Sections for the Elliptic Umbilic Caustic Surface}

\author{Amir B.\ Aazami}
\affiliation{Department of Mathematics \& Computer Science,\\ Clark University,\\
  Worcester, MA 01610;\\
  {\tt aaazami@clarku.edu}}

\author{Charles R.\ Keeton}
\affiliation{Department of Physics \& Astronomy, Rutgers University,
  136 Frelinghuysen Road, Piscataway, NJ 08854-8019;\\
  {\tt keeton@physics.rutgers.edu}}

\author{Arlie\ O.\ Petters}
\affiliation{Departments of Mathematics and Physics, Duke University,\\
  Science Drive, Durham, NC 27708-0320;\\
  {\tt arlie.petters@duke.edu}}


\begin{abstract}
\noindent In gravitational lensing, magnification cross sections characterize the probability that a light source will have magnification greater than some fixed value, which is useful in a variety of applications. The (area) cross section is known to scale as $\mu^{-2}$ for fold caustics and $\mu^{-2.5}$ for cusp caustics. We aim to extend these results to higher-order caustic singularities, focusing on the elliptic umbilic, which can be manifested in lensing systems with two or three galaxies. The elliptic umbilic has a caustic surface, and we show that the volume cross section scales as $\fkM^{-2.5}$ in the two-image region and $\fkM^{-2}$ in the four-image region, where $\fkM$ is the total unsigned magnification. In both cases our results are supported both numerically and analytically.
\end{abstract}
\maketitle

\section{Introduction}
\label{Introduction}
Magnification cross sections are an important tool in gravitational lensing.  Knowing the magnification cross section allows one to determine the probability that a light source will have magnification greater than some fixed value.  This in turn gives information about the accuracy of cosmological models, since these predict different probabilities regarding source magnifications (see Schneider et al. 1992~\cite{Sch-EF}, Kaiser 1992~\cite{Kaiser}, Bartelmann et al.\ 1998~\cite{Bart}, and Petters et al.\ 2001~\cite[Chapter~13]{Petters}).  Knowing magnification cross sections is also important for observational programs that use lensing magnification to help detect extremely faint galaxies (see, e.g., Lotz et al. 2017~\cite{Lotz}, and references therein).

It is well known that for the so-called ``fold'' and ``cusp'' caustic singularities, the area cross sections scale asymptotically as $\fkM^{-2}$ and $\fkM^{-2.5}$, respectively, where $\fkM$ is the total unsigned magnification of a lensed source (for the cusp caustic, it is assumed that the source lies in the one-image region locally).  In the case of single-plane lensing, the fold scaling was determined by Blandford and Narayan 1986~\cite{Blan-Nar}, while the cusp scaling was determined by Mao 1992~\cite{Mao} and Schneider and Weiss 1992~\cite{Sch-Weiss92}.  For multiple-plane lensing, the fold and cusp scalings were determined by Petters et al.~\cite[Chapter 13]{Petters}.

In this paper we commence the study of magnification cross sections of ``higher-order'' caustic surfaces; in particular, we derive the (single-plane) asymptotic limit of the magnification cross section for the ``elliptic umbilic'' caustic surface, which is not a curve but rather a two-dimensional stable caustic surface in a three-dimensional parameter space (for precise definitions, consult, e.g., Arnold 1973~\cite{Arnold73}, Callahan 1974 \& 1977~\cite{Callahan, Callahan2}, Majthay 1985~\cite{Majthay}, Arnold et al. 1985~\cite{AGV1}, Petters 1993~\cite{Petters93}, \cite{Sch-EF,Petters}); its magnification ``cross section'' is therefore a region with volume.  As hypothesized in Rusin et al. 2001~\cite{Rusin} and Blandford 2001~\cite{Blandford2001}, the elliptic umbilic is likely to be manifested inside a triangle formed by three lensing galaxies, and to involve one positive-parity and three negative-parity images.  As shown in Shin and Evans 2008~\cite{Shin-Evans} and de Xivry and Marshall 2009~\cite{O-Marshall}, elliptic umbilics can also appear in lensing by binary galaxies, if the binary separation is small enough. The asymptotic scaling of the elliptic umbilic volume cross section will depend on whether the source gives rise to two or four lensed images locally. We show that, in the two-image region, this volume cross section scales to leading order as $\fkM^{-2.5}$, whereas in the four-image region, its leading order scales as $\fkM^{-2}$.  In both cases, our results are supported numerically and analytically. In our derivation of $\fkM^{-2}$ in the four-image region, we make use of a certain magnification relation that holds for higher-order caustic singularities, and the elliptic umbilic in particular, that was shown to hold in Aazami and Petters 2009~\cite{AP}.

\section{The elliptic umbilic caustic surface}
\begin{figure}[t]
\begin{center}
\includegraphics[scale=.6]{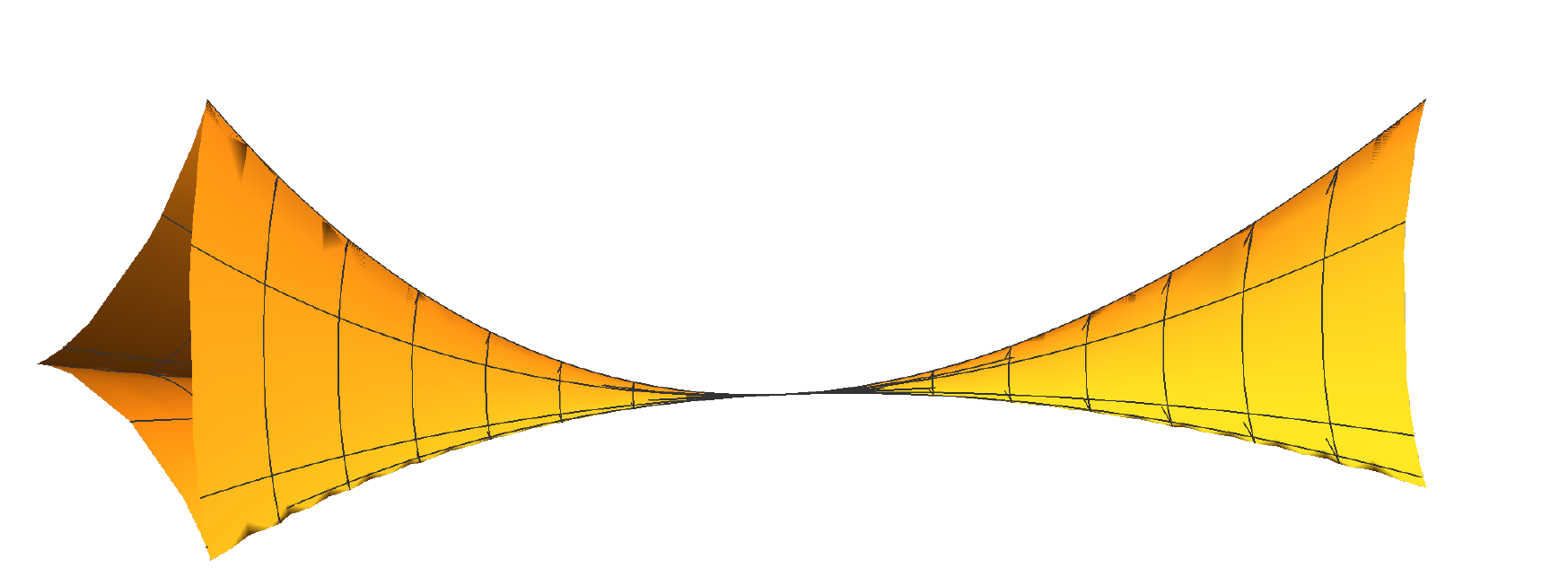}
\end{center}
\caption{The elliptic umbilic caustic surface in three-dimensional parameter space $\{(s_1,s_2,c)\} = \RR^3$; $s_1,s_2$ are source plane coordinates and $c \in \RR$ an additional parameter.  When $c = 0$ the elliptic umbilic is the central point shown. A $c$-slice of this caustic is shown in Figure \ref{figure2}.}
\label{figure1}
\end{figure}

\begin{figure}[t]
\begin{center}
\includegraphics[scale=.49]{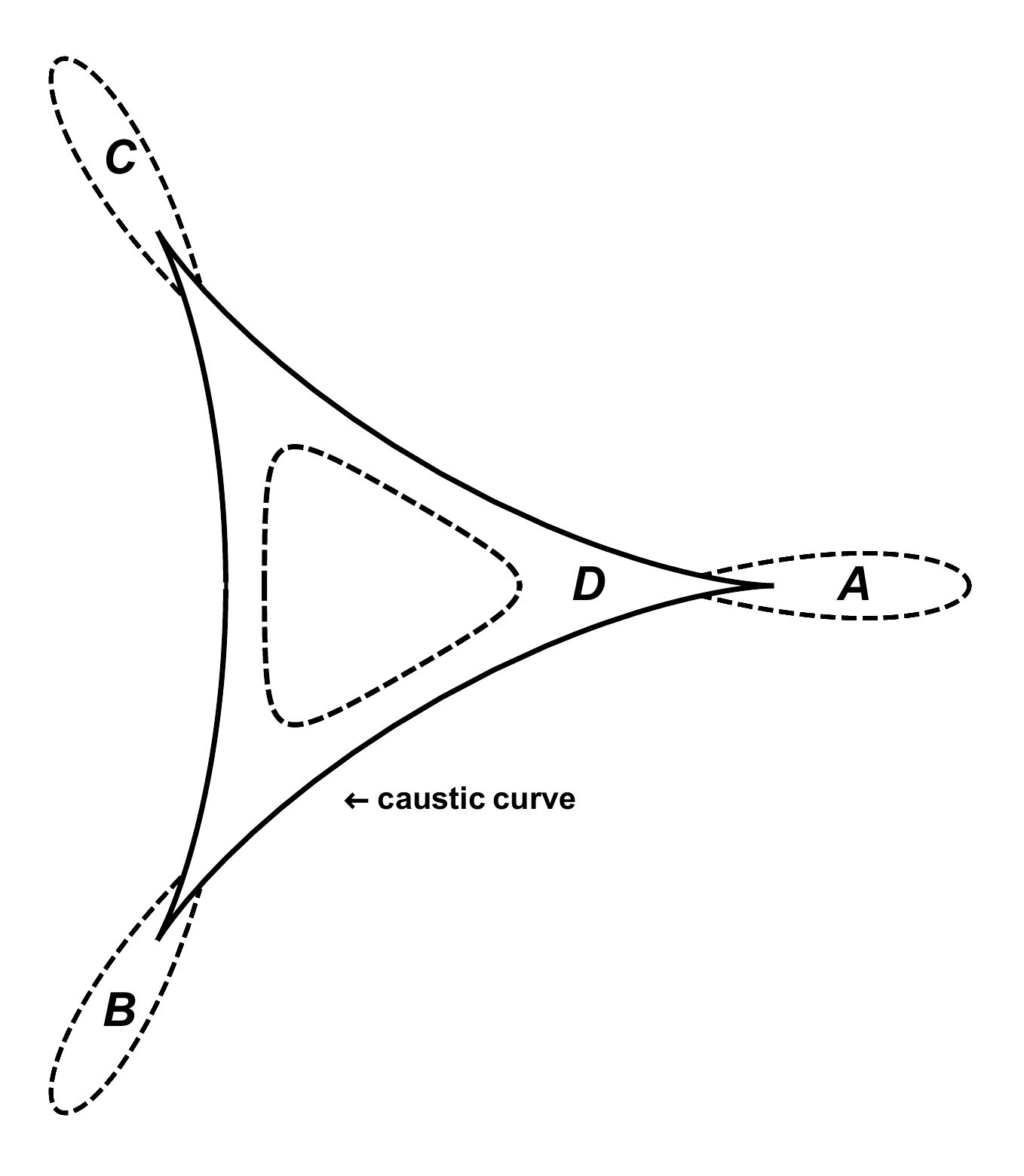}
\end{center}
\caption{A generic $c$-slice of the elliptic umbilic caustic surface, which is the triangular-shaped solid curve. For level curves in the two-image region, which is the (unbounded) region outside the closed caustic curve, the magnification volume cross section $\sigma(\mu)$ is, for $\mu$ large enough, the sum of the closed regions $A,B,C$.  A source inside regions $A,B$, or $C$ has two lensed images and total unsigned magnification greater than $\mu$.  For level curves in the four-image region, which is inside the triangular-shaped solid curve, $\sigma(\mu)$ is, for $\mu$ large enough, the region $D$ inside the triangular-shaped caustic but outside the closed dashed curve within.  A source in region $D$ has four lensed images and total unsigned magnification greater than $\mu$. Note that the actual caustic and level ``curve'' are both surfaces in $\RR^3$; what is shown here is a $c$-slice of this surface on the source plane.}
\label{figure2}
\end{figure}

Let $(x,y)$ denote coordinates on the lens plane and $(s_1,s_2)$ those on the source plane.  Like all higher-order caustic surfaces, the \emph{elliptic umbilic caustic} has parameters in addition to the two source plane coordinates.  (Such higher-order parameters can, depending on the setting, be used to model the source redshift, radii of galaxies, ellipticities, distance along the line of sight, etc., of the lens system in question; see, e.g., \cite[Chapter 8]{Sch-EF}). A gravitational lensing map in the neighborhood of an elliptic umbilic critical point, as derived in \cite[Chapter 5]{Sch-EF}, takes the form
\beqa
\label{eqn:ell}
\eta(x,y) = (x^2-y^2,-2xy+4cy) = (s_1,s_2),
\eeqa
where $c \in \RR$ is a parameter in addition to the source plane coordinates $s_1,s_2$; i.e., the elliptic umbilic caustic is a surface in the parameter space $\{(s_1,s_2,c)\} = \RR^3$, not a curve in the source plane $S := \{(s_1,s_2)\} = \RR^2$. Accordingly, the ``magnification cross section" is a (three-dimensional) volume, and the asymptotic scaling is the leading order term in the limit as the magnification goes to infinity.  Figure~\ref{figure1} shows the elliptic umbilic caustic surface, while Figure~\ref{figure2} shows a $c$-slice of it on the source plane $S$.

If a source located at $(s_1,s_2)$ on the source plane has a lensed image located at $(x,y)$ on the lens plane, then the magnification $\tilde{\mu}$ of this lensed image is given by
\beq
\label{hyp}
\tilde{\mu}(x,y) = \frac{1}{{\rm det(Jac}\,\eta)(x,y)} = \frac{1}{8c x-4(x^2+y^2)}\cdot
\eeq

Now fix $c \in \RR$ and $\mu > 0$.  Consider first the four-image region enclosed by the caustic curve in Figure \ref{figure2}. Let $C_\mu$ denote the subset consisting of those source positions $(s_1,s_2)$ in the four-image region with total \emph{unsigned} magnification equal to $\mu$, where the unsigned magnification of each image $(x_i,y_i)$ belonging to $(s_1,s_2)$ is given by $|\tilde{\mu}(x_i,y_i)|$ in \eqref{hyp}. $C_\mu$ will consist of the closed dashed curve inside the caustic curve in Figure \ref{figure2}; any source \emph{inside} the dashed region will have total unsigned magnification \emph{less} than $\mu$, while a source in the region $D$ will have total unsigned magnification equal to $\mu$.  Likewise for the two-image region, which is the (unbounded) region outside the caustic curve: for $\mu$ large enough, the enclosed regions $A$, $B$, and $C$ comprise those sources whose two lensed images will have total unsigned magnification greater than $\mu$. (By symmetry, $A$, $B$, and $C$ all have the same areas.)

Whether in the two- or the four-image region, the asymptotic scaling there is determined as follows.  The areas labeled $A$, $B$, $C$, and $D$ will in general be functions of $\mu$ and $c$; denote any one of these, e.g., by $A(\mu,c)$.  To obtain a \emph{volume} section, denoted $V(\mu)$, we integrate
\beq\label{eqn:V}
V(\mu) = \int_{c_1}^{c_2} A(\mu,c)\,dc,
\eeq
where $c_1 < c_2$ are arbitrary but of the same sign. Finally, we take the limit
$
\lim_{\mu\to \infty}V(\mu)
$
and identify the leading order term; this is the asymptotic scaling of the elliptic umbilic magnification volume.
\begin{figure}[t]
\begin{center}
\includegraphics[width=0.7\textwidth]{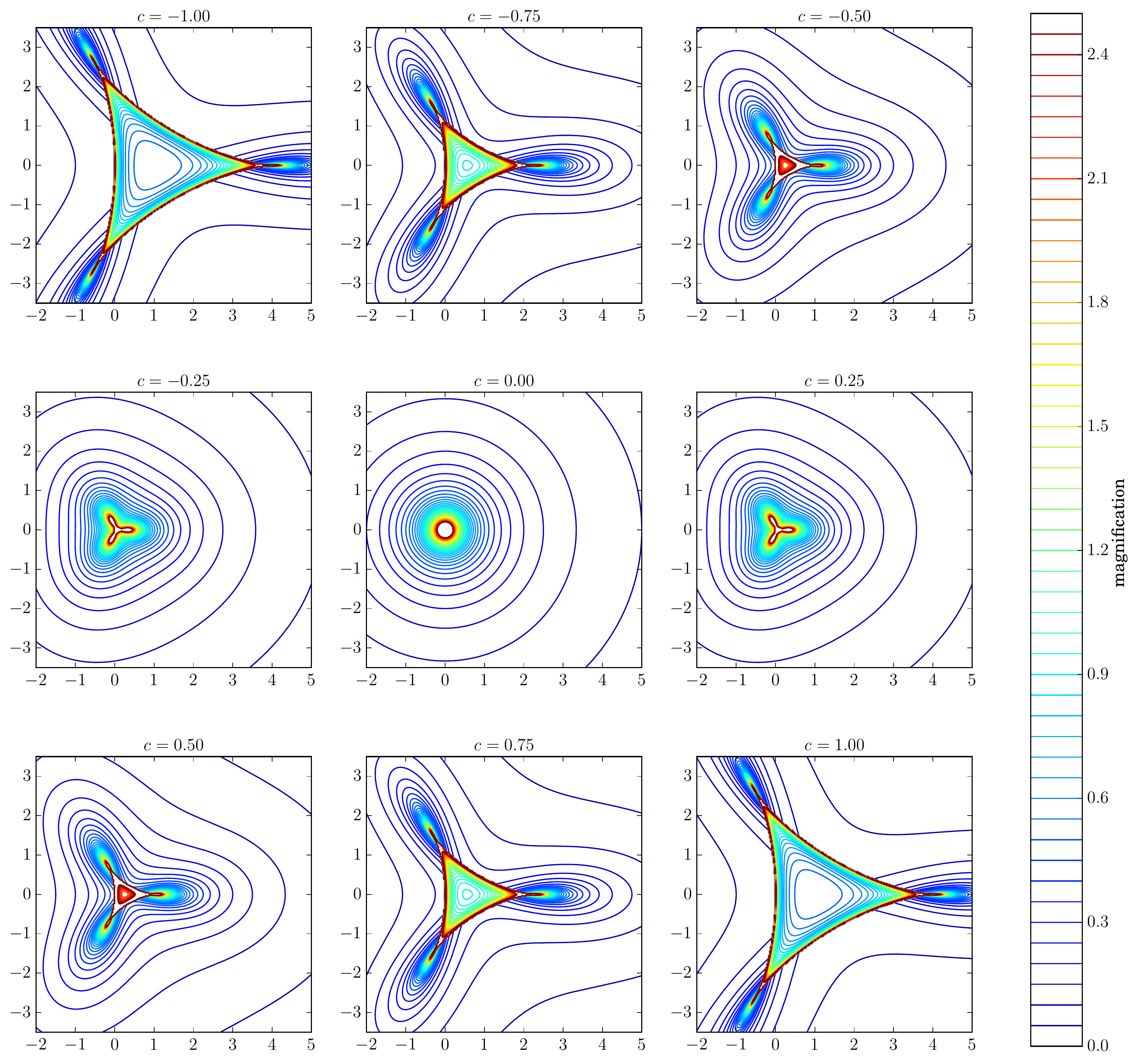}
\end{center}
\caption{
Level curves of the total unsigned magnification, for different values of $c$.
}
\label{numerical_levelcurves}
\end{figure}

\begin{figure}[t]
\begin{center}
\includegraphics[scale=.6]{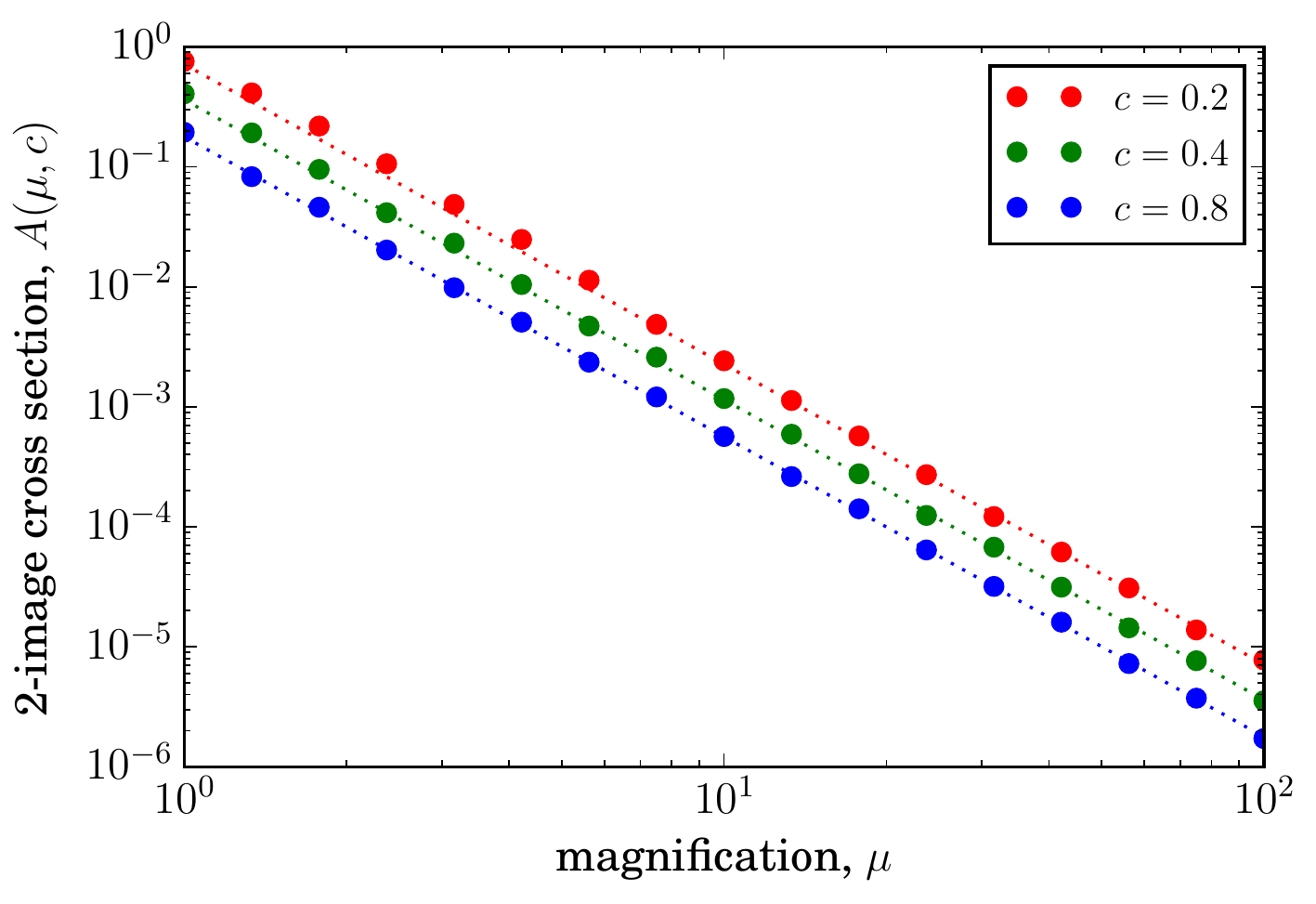}
\includegraphics[scale=.6]{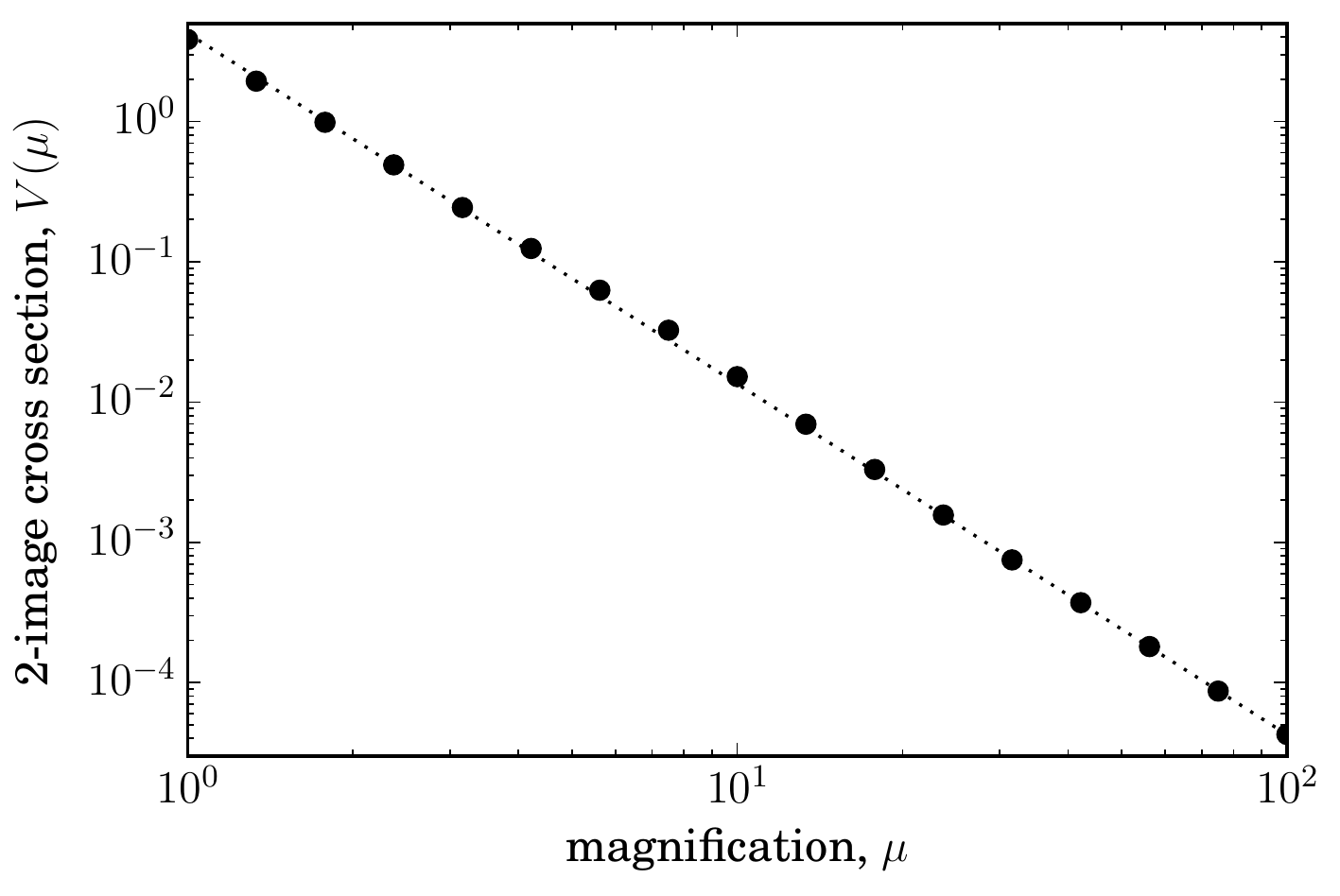}
\end{center}
\caption{
\emph{(Left)} The colored points show the area cross section $A(\mu,c)$ in the two-image region, computed numerically for different $c$-slices.  The dotted lines show the scaling $A(\mu,c) \propto \mu^{-2.5}$.
\emph{(Right)} The points show the volume cross section $V(\mu)$ in the two-image region, computed by integrating $c$ over the range $[0.1,1.0]$.  The dotted line shows the scaling $V(\mu) \propto \mu^{-2.5}$.
}\label{numerical_csec2}
\end{figure}

\begin{figure}[t]
\begin{center}
\includegraphics[scale=.6]{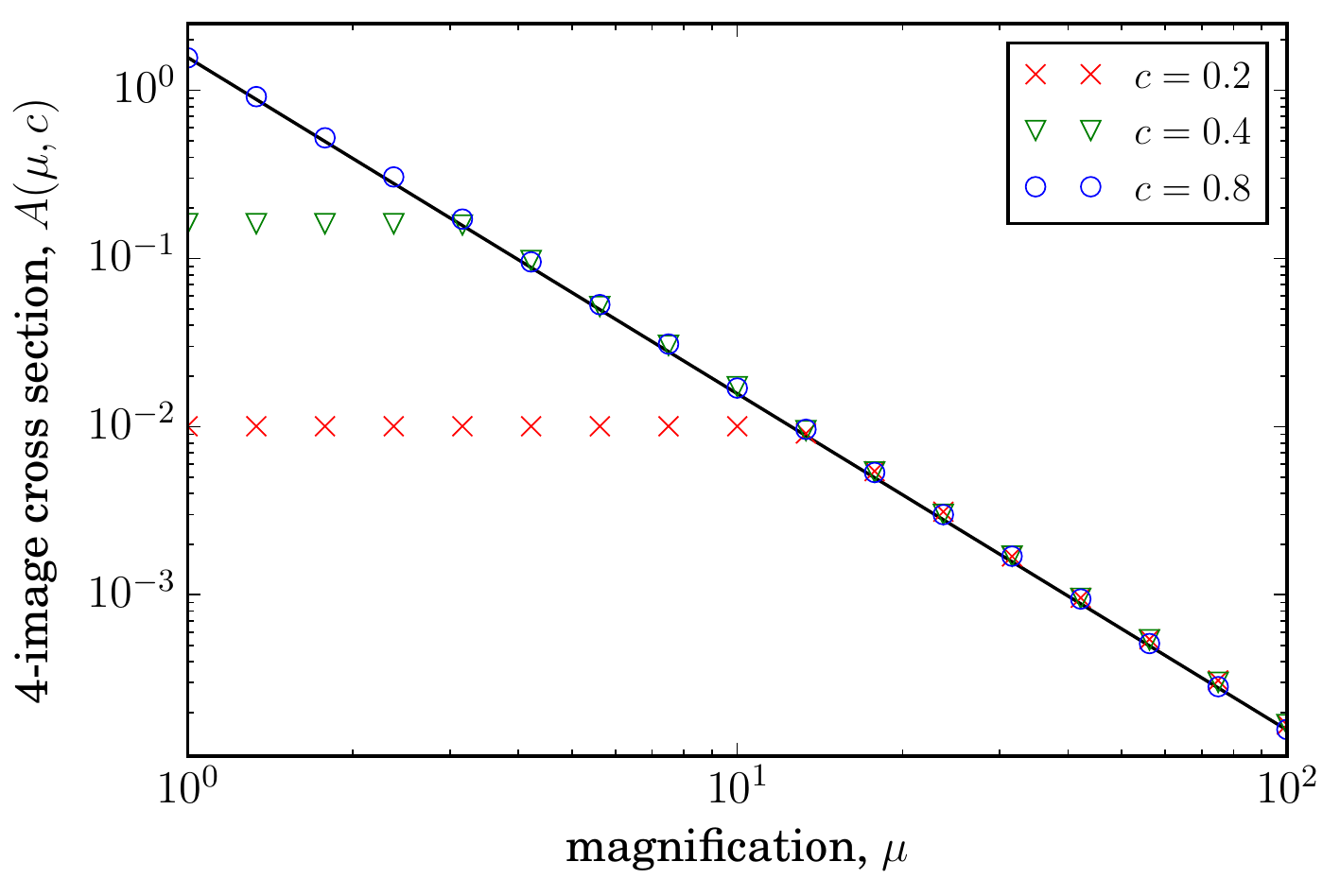}
\end{center}
\caption{
The colored points show the area cross section $A(\mu,c)$ in the four-image region, computed numerically for different $c$-slices.  The solid line shows $A(\mu,c) = \pi/(2\mu^2)$, in agreement with the result obtained analytically in \eqref{eqn:4scal} (with $\tilde{\mu} = \mu/2$). The cross section curves level off at the magnification that corresponds to the center of the caustic.
}\label{numerical_csec4}
\end{figure}

We first examine the scalings numerically.  For a given $c$-slice, we compute the total magnification on a grid in plane $S$, as shown in Figure \ref{numerical_levelcurves}.  The grid is adaptive, meaning that more grid points are used in regions where the magnification changes quickly and high resolution is needed to obtain accurate results.  We use the grid to approximate the area integral and compute $A(\mu,c)$.  We then combine different $c$-slices to approximate the integral in \eqref{eqn:V} and obtain $V(\mu)$.  Figure \ref{numerical_csec2} shows examples of $A(\mu,c)$ and $V(\mu)$ for the two-image region, while Figure \ref{numerical_csec4} shows $A(\mu,c)$ for the four-image region.

We support our numerical findings with the following analytical arguments.  For the four-image region, there is in fact a succinct argument that confirms the numerical result $\mu^{-2}$, as follows. For any source in the four-image region (and assuming $c \neq 0$), the lensing map \eqref{eqn:ell} has three lensed images with negative magnification and one lensed image with positive magnification, where the magnification of a lensed image is given by \eqref{hyp}. Fix $\tilde{\mu} > 0$ and let $S_{\tilde{\mu}}$ denote the set of sources in the four-image region whose one positive-magnification image has magnification $\tilde{\mu}$; as usual, let $\mu$ denote the total unsigned magnification of this source. In fact $S_{\tilde{\mu}}$ comprises the closed dashed curve inside the caustic that we saw in Figure \ref{figure2}.  It is straightforward to compute that the area labeled $D$\,---\,that is, the area outside $S_{\tilde{\mu}}$ but inside the caustic\,---\,is equal to
\beqa
\label{eqn:4scal}
2\pi c^4 - \Big(2\pi c^4 - \frac{\pi}{8\tilde{\mu}^2}\Big)  = \frac{\pi}{8\tilde{\mu}^2}\cdot
\eeqa
Observe that this area is $c$-independent.  This area is related to $A(\mu,c)$, the magnification (area) cross section, as follows.  Let $\vec{s}$ be a source in the four-image region whose total unsigned magnification is $\mu$.  Clearly $\vec{s} \notin S_\mu$, since its one positive-magnification image must be less than $\mu$.  However, it is known that the four-image region of the elliptic umbilic, for any $c \neq 0$, satisfies the following magnification relation,
$$
\sum_{i=1}^4 \tilde{\mu}_i = 0,
$$
with $\tilde{\mu}_i$ is the signed magnification of lensed image $i$; see \cite{AP} for a proof. (For clarity, we use the notation ``$\tilde{\mu}$" to denote the magnification belonging to an individual image, and reserve the notation ``$\mu$" to denote the total unsigned magnification of a source.) It follows that $\vec{s} \in S_{\mu/2}$, because the positive-magnification image must have magnification $\mu/2$, since the other three magnifications are negative and must cancel it out.  Thus the closed curve $S_{\mu/2}$ is precisely the level curve of sources with total unsigned magnification $\mu$.  Observe that when $\tilde{\mu} = \mu/2$ in \eqref{eqn:4scal}, then $A(\mu,c) = \pi/(2\mu^2)$, as in Figure \ref{numerical_csec4}. And thus the area cross section scales like \eqref{eqn:4scal}.  Figure \ref{numerical_csec4} confirms this result numerically.  Integrating \eqref{eqn:4scal} from $c_1$ to $c_2$ yields a volume cross section that clearly scales as $V(\mu) \propto  \mu^{-2}$.

\begin{figure}[t]
\begin{center}
\includegraphics[scale=.3]{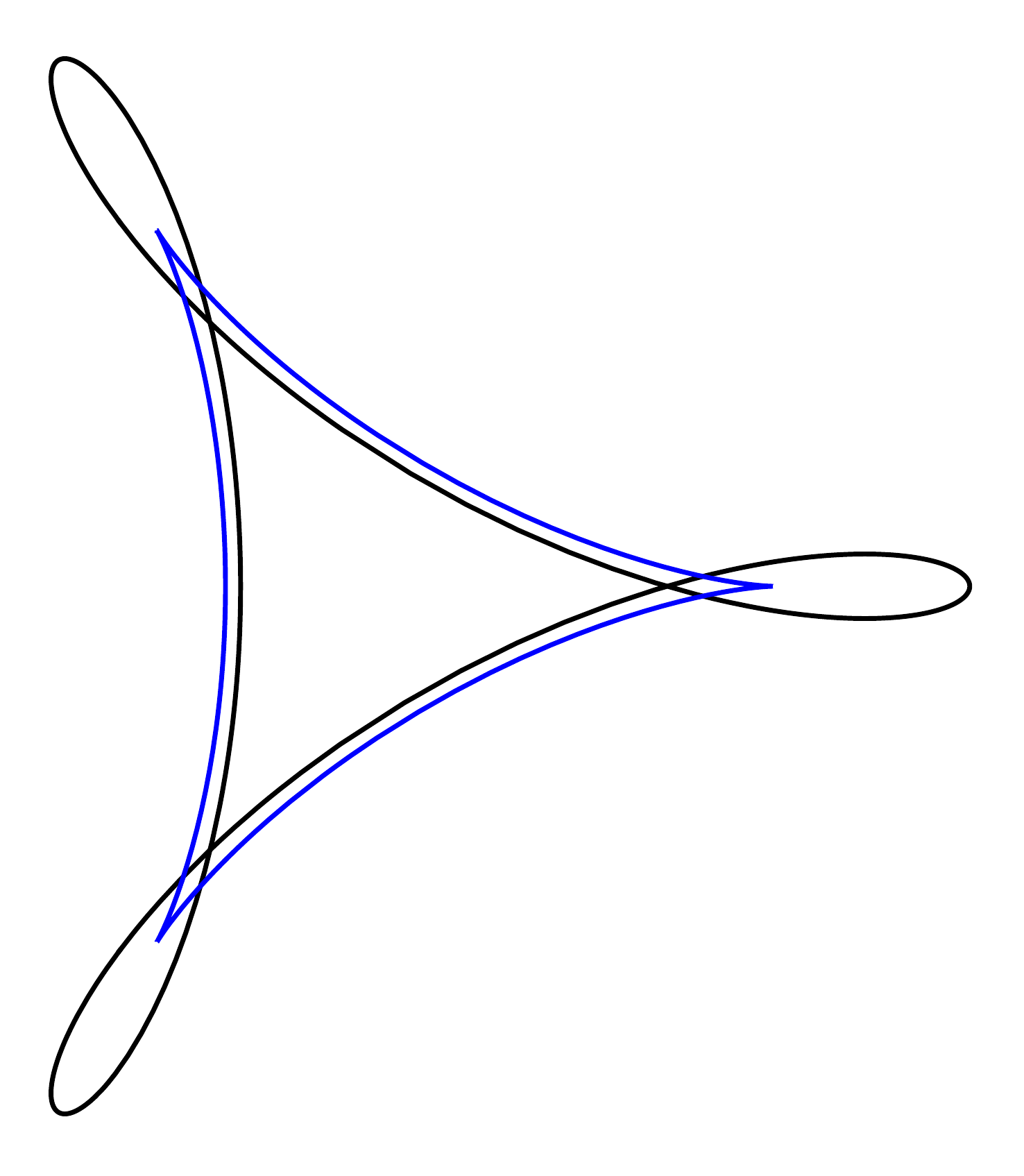}
\includegraphics[scale=.265]{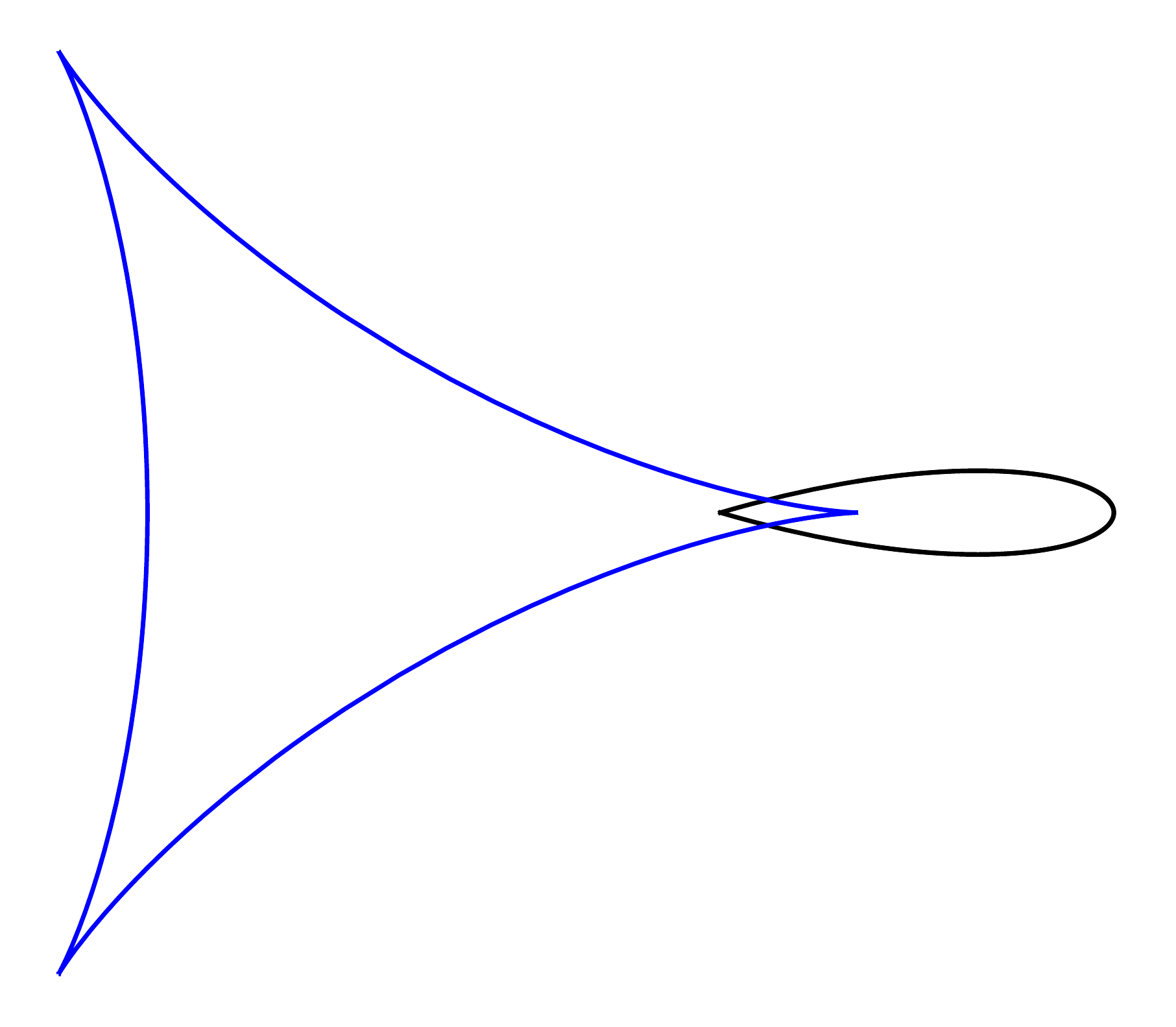}
\includegraphics[scale=.3]{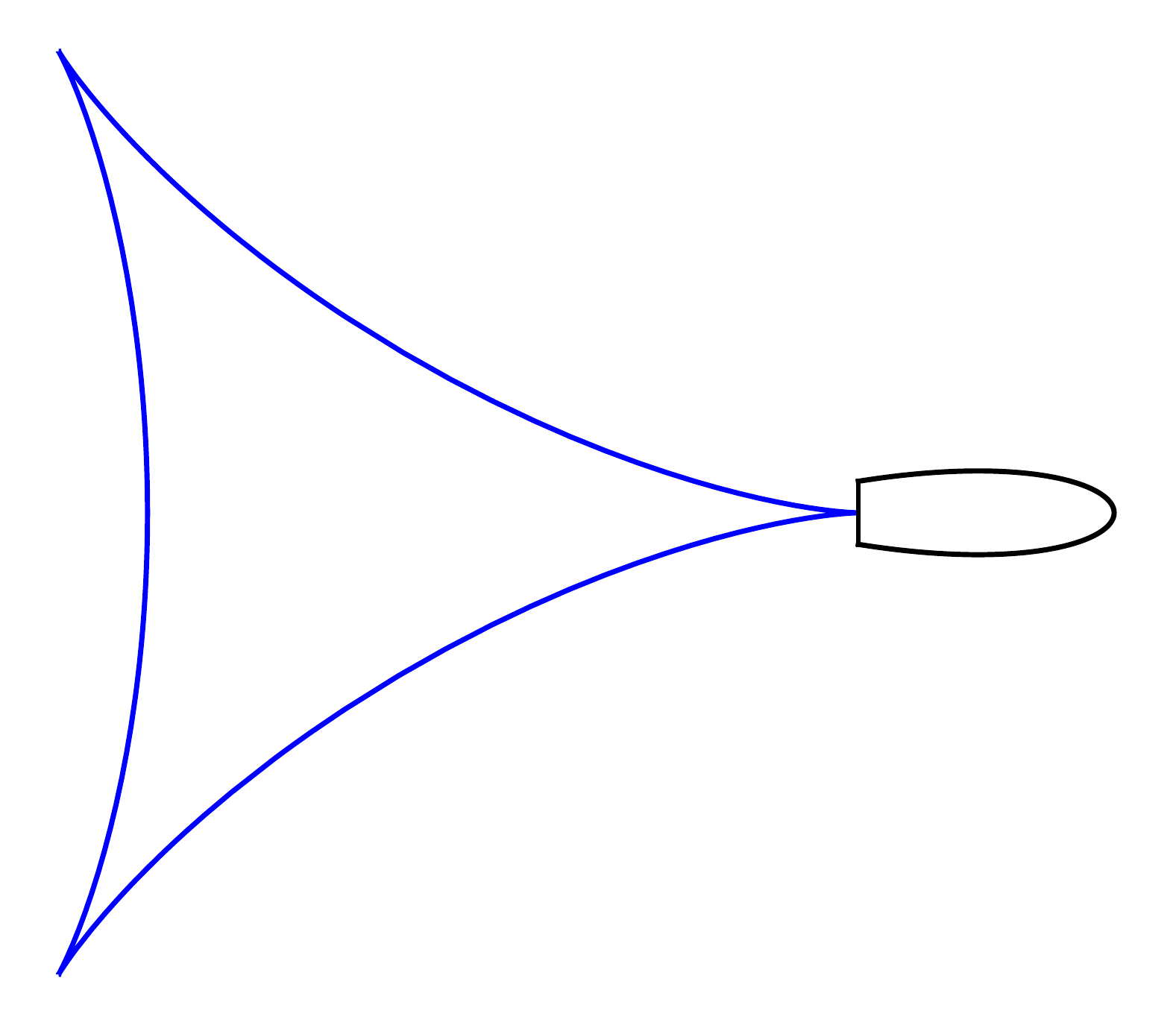}
\end{center}
\caption{In each panel, the blue triangular-shaped curve is (a $c$-slice of) the elliptic umbilic caustic.  In the leftmost figure, the black curve comprises all source positions with at least one image with (unsigned) magnification $\tilde{\mu}$, for some fixed value of $\tilde{\mu}$. In the large magnification limit, the magnification of one of the images in the two-image region dominates that of the other, in which case the loops enclosing each cusp in the leftmost panel approach the regions $A,B,C$ in Figure \ref{figure2}. The closed black curves in the middle and rightmost figures bound the areas $A,B$, and $C$ from above and below. It can be shown that these two area bounds, which are functions of $\mu$ and $c$ in general, both scale to leading order in $\mu$ as $\sim\!\mu^{-2.5}$, where $\mu$ is the total unsigned magnification of a source in the two-image region.
}
\label{fig:6}
\end{figure}

There remains, finally, the two-image region outside the caustic curve shown in Figure \ref{figure2}. Here, for $\mu > 0$ large enough, those sources with total unsigned magnification \emph{greater} than $\mu$ comprise the regions labeled $A$, $B$, and $C$, which three enclosed areas are equal by symmetry. In this case the area enclosed by them is not as easily derivable analytically, due to the complicated nature of the intersection points of the dashed curves with the caustic. However, it is possible to \emph{bound} the areas $A$, $B$, and $C$, from above and below, and to show that these lower and upper bounds, which are functions of $\mu$ and $c$, both scale to leading order in $\mu$ as $\sim\!\mu^{-2.5}$, thereby supporting our numerical results in Figure \ref{numerical_csec2}. Figure \ref{fig:6} briefly describes the procedure, foregoing technical details.

\section*{CONCLUSION}
We have shown that the asymptotic scaling of the magnification volume cross section corresponding to an elliptic umbilic caustic surface is $\fkM^{-2.5}$ in the two-image region and $\fkM^{-2}$ in the four-image region, where $\fkM$ is the total unsigned magnification. In both cases our results are supported both numerically and analytically. The goal was to extend to higher-order caustic singularities the well known (area) cross sections for fold ($\mu^{-2}$) and cusp ($\mu^{-2.5}$) caustics, in particular given that the elliptic umbilic caustic can be manifested in both binary and three-galaxy lensing systems.

\section*{ACKNOWLEDGMENTS}
\noindent Some of this work was done at the American Mathematical Society's June, 2018, Mathematics Research Conference on the ``Mathematics of Gravity and Light."  This material is based upon work supported in part by the National Science Foundation under Grant No. 1641020. 

\bibliographystyle{mdpi}

\end{document}